\documentclass[aps,prr,amsmath,amssymb,floatfix,twocolumn,10pt]{revtex4-1}
\usepackage{graphicx}
\usepackage{subfigure}
\usepackage{hyperref}
\usepackage{braket}
\usepackage{bm}
\usepackage{comment}
\usepackage{xcolor}
\usepackage{soul}
\setstcolor{red}

\DeclareMathAlphabet\mathbfcal{OMS}{cmsy}{b}{n}

\def\YBCOs{YBa$_2$Cu$_3$O$_{6+x}$ }
\def\YBCOns{YBa$_2$Cu$_3$O$_{6+x}$}
\def\YBCO248{YBa$_2$Cu$_4$O$_8$}

\def\LBCO{La$_{2-x}$Ba$_{x}$CuO$_4$ }

\def\ie{{\it i.e.}}

\def\bPhi{{\bm \Phi}}
\def\bnabla{\bm{\nabla}}
\def\rhos{\rho_s}
\def\D{\mathcal{D}}

\def\br{{\bf r}}

\def\bk{{\bf k}}
\def\bq{{\bf q}}
\def\bQ{{\bf Q}}

\def\bQ{{\bf Q}}

\def\dr{{d^2r}}
\def\Tr{{\rm Tr}}

\def\tU{\tilde{U}}
\def\tJ{\tilde{J}}
\def\cF{{\cal F}}
\def\cL{{\cal L}}
\def\prl{{Phys. Rev. Lett. }}
\def\prb{{Phys. Rev. B }}

\def\science{{Science }}
\def\nature{{Nature (London) }}
\def\natphys{{Nat. Phys. }}
\def\natcomm{{Nat. Commun. }}
\def\cV{{\cal V}}
\def\cG{{\cal G}}

\def\tW{\tilde{W}}

\newcommand\redsout{\bgroup\markoverwith{\textcolor{red}{\rule[0.5ex]{2pt}{0.4pt}}}\ULon}

\makeatletter
\newsavebox{\@brx}
\newcommand{\llangle}[1][]{\savebox{\@brx}{\(\m@th{#1\langle}\)}%
  \mathopen{\copy\@brx\kern-0.5\wd\@brx\usebox{\@brx}}}
\newcommand{\rrangle}[1][]{\savebox{\@brx}{\(\m@th{#1\rangle}\)}%
  \mathclose{\copy\@brx\kern-0.5\wd\@brx\usebox{\@brx}}}
\makeatother

\begin{document}

\title{Disorder effects in a model of competing superconducting and charge-density wave orders in YBa$_2$Cu$_3$O$_{6+x}$}

\author{Dror~Orgad}
\affiliation{Racah Institute of Physics, The Hebrew University,
  Jerusalem 91904, Israel}

\date{\today}

\begin{abstract}
There is evidence for competition between superconductivity and short-range charge-density wave order
in a number of cuprate compounds and especially in YBa$_2$Cu$_3$O$_{6+x}$. Here, we use a non-linear sigma model of such
competition to study the effects of spatially correlated disorder in the chain layers and delta-correlated disorder in the CuO$_2$
layers. The first is relevant to the oxygen ortho structure and the latter may be induced by electron irradiation.
We find that reducing the correlation length of the potential on the chain layers decreases the size of the
charge-density wave structure factor but has little effect on its temperature dependence, as observed experimentally.
At the same time, the charge-density-wave correlation length decreases more than is seen in the experiment. The temperature
at which the structure factor peaks coincides with $T_c$ and both decrease when disorder is introduced into the CuO$_2$ planes.
Strengthening this disorder reduces the magnitude of the structure factor and eventually turns it into a monotonic function
of the temperature. This occurs despite an increase in the local magnitude of the charge-density wave order.
\end{abstract}

\maketitle

\section{Introduction}

Signatures of short-range charge-density wave (CDW) order have been detected in a wide range of cuprate superconductors
\cite{Tranquada95,Zimmermann98,Hunt99,Hucker11,Ghiringhelli,Chang,Achkar1,Blackburn13,Comin1,daSilva,Le-Tacon,Hucker14,
Blanco-Canosa14,Croft,Tabis,SilvaNeto,Comin3,Comin-symmetry,Forgan-structure,NMR-shortcor,Peng-Bi2201,Tabis2017}.
The short-range CDW appears to compete with superconductivity and in \YBCOs (YBCO) it is possible to stabilize a much longer-ranged
order with the same incommensurate wavelength by suppressing superconductivity using either strong magnetic fields,
\cite{NMR-nature,NMR-nature-comm,ultrasound,Gerber3D,Chang3D,Jang3D,Chang3D-2020} or strain
\cite{Kim-strain-science,Kim-strain-PRL,Vinograd24}. Another agent that is capable of tipping the balance between the
two ordering tendencies is the disorder present in any physical sample.

Previous works have introduced Ginzburg-Landau and nonlinear sigma models (NLSM) to study the competition between superconducting
and CDW orders in the absence \cite{Zachar,Demler,Efetov,Hayward1} and presence of disorder \cite{Nie,NLSM-layers,NLSM-bilayer,Vinograd24}.
In particular, we have shown that increasing the strength of a delta-correlated Gaussian disorder 
leads to larger short-range CDW correlations at low temperatures \cite{NLSM-layers}. If interpreted as
a universal relation between disorder strength and CDW correlations this appears to conflict with
an experiment by Achkar {\it et al}. \cite{Achkar-disorder}, who used quench cooling to destroy the ortho structure
of the doped oxygens in the YBCO chain layers \cite{Andersen-ortho} and resonant x-ray scattering to
reveal that such disordering of the dopant atoms leads to uniform {\it reduction} of the CDW structure factor across all temperatures.

Our studies of an NLSM adopted to reflect the structure of YBCO have demonstrated its ability to reproduce many of the salient
features of the observed CDW signatures in this compound under the influence of magnetic fields \cite{NLSM-bilayer} and strain \cite{Vinograd24}.
However, while these works included a random potential on the chain layers, we have not yet investigated the effects of varying the
correlation length of the potential. Hence, it remains uncertain whether the same NLSM can also capture the observed reduction in
the CDW correlations due to oxygen disordering. In the following we answer this question in the affirmative using numerical Monte Carlo
simulations.

Specifically, we show that increasing the correlation length of the potential in the chain layers induces an increase in the CDW structure
factor, even when the variance of the random potential is simultaneously reduced. This indicates that a correlated chain
potential is more effective in nucleating CDW order than a local Gaussian disorder. Furthermore, the relative increase in the CDW
structure factor is largely independent of the temperature, as was observed experimentally \cite{Achkar-disorder}.
Similar statements hold for the magnitude and temperature dependence of the calculated CDW correlation length. Here, however,
we find a discrepancy with the experimental results \cite{Achkar-disorder} that exhibit considerably smaller changes in the
magnitude of the CDW correlation length following the disordering of the ortho structure. Given the overall success of the model
in reproducing the observed phenomenology, I take this discrepancy as a guidance for further theoretical refinements.

In YBCO, disorder naturally arises from oxygen doping and is mostly confined to the chain layers. However,
it is possible to introduce point defects directly into the CuO$_2$ planes by electron irradiation \cite{Alloul-irradiation}.
This raises the intriguing question of how such disorder influences the interplay between the CDW correlations and superconductivity.
We have not addressed this issue before within our model, nor are we familiar with relevant experiments on YBCO.
To fill the gap and provide a theoretical reference point for future experiments we describe below
our findings regarding the behavior of the NLSM with additional Gaussian disorder on the CuO$_2$ planes.
We show that although the additional disorder {\it enhances} the amplitude of the local CDW order it causes a {\it reduction}
in the size of the CDW structure factor owing to a decrease in the CDW correlation length. It also affects the temperature
dependence of the structure factor, which for not too strong disorder exhibits a peak at the superconducting $T_c$. Increasing
the disorder reduces $T_c$ and eventually turns the structure factor into a monotonically decreasing function of the temperature.

Finally, and from a more analytical perspective, we aim to explore the extent to which the properties of the disordered
NLSM can be accurately captured by the large-$N$ approximation ($N$ being the number of components of the vector of order parameters).
We show that while the saddle-point equations (infinite-$N$ limit) reproduce the broad qualitative features of the physical model,
they miss some of the quantitative aspects. Including $1/N$ corrections into the analysis improves the quantitative agreement
at high temperature but introduces increased deviations in the vicinity of $T_c$.

\section{The Model}

We consider a model consisting of $L_z/2$ bilayers representing the CuO$_2$ bilayers of YBCO. Each layer hosts a
three-dimensional complex order parameter whose components correspond to a superconducting order parameter $\psi_{j\mu}(\br)$,
and two complex CDW order parameters $\Phi^{a,b}_{j\mu}(\br)$. The latter describe density variations
$\delta\rho_{j\mu}(\br)=e^{i\bQ_a\cdot\br}\Phi^a_{j\mu}(\br)+e^{i\bQ_b\cdot\br}\Phi^b_{j\mu}(\br)+{\rm c.c.}$ along the
crystallographic $a$ and $b$ directions with incommensurate wave-vectors $\bQ_{a,b}$. Here,
$\mu=0,1$ denotes the bottom (top) layer within a bilayer, which by itself is indexed by $j$.
In the following we coarse grain the planes, such that the in-plane position vector $\br$ corresponds to sites on a square
lattice whose lattice constant $a$ is the observed CDW period, i.e., about three Cu-Cu spacings.

We focus on temperatures below the onset temperature of short-range CDW correlations (135-155K in YBCO) and assume
the existence of some type of local order at every lattice point. The competition between the different components is
encapsulated by the constraints
\begin{equation}
\label{meq:constraint}
|\psi_{j\mu}(\br)|^2+|\bPhi_{j\mu}(\br)|^2=1,
\end{equation}
where ${\bm \Phi}_{j\mu}=(\Phi_{j\mu}^a,\Phi_{j\mu}^b)^T$.
The Hamiltonian is
\begin{eqnarray}
\label{meq:HV}
\nonumber
H&=&\sum_j\sum_{\mu=0,1} H_{j\mu}+\frac{\rhos}{2}\sum_j\sum_\br
\Big[\tU \bPhi^\dagger_{j,0}\bPhi_{j,1}\\
\nonumber
&&+U\bPhi^\dagger_{j,1}\bPhi_{j+1,0}
-\tJ\psi_{j,0}^*\psi_{j,1}-J\psi_{j,1}^*\psi_{j+1,0}\\
&&+{\bm V}^\dagger_j\left(\bPhi_{j,1}+\bPhi_{j+1,0}\right)+{\rm c.c.}\Big].
\end{eqnarray}
Henceforth, the bare superconducting stiffness, $\rhos$, and the lattice constant $a$ are set to 1 and serve as the basic
energy and length scales.
We model the Coulomb interaction between CDW fields within a bilayer by a local coupling $\tU$, and denote
the intra-bilayer Josephson tunneling amplitude by $\tJ$.
The (weaker) Coulomb interaction and Josephson coupling between nearest-neighbor planes belonging to
consecutive bilayers are denoted by $U$ and $J$, respectively.

The last term in Eq. (\ref{meq:HV}) describes the coupling between the disordered doped oxygens on the chain layers and the
CDW fields on the adjacent bilayers. This coupling may originate from local changes in the concentration of doped holes
and from the Coulomb interaction between the oxygens and the CDW \cite{Tabis2017}. We model the chain potential via
correlated random fields ${\bm V}_j=(V^1_j + iV^2_j, V^3_j + iV^4_j)^T$ satisfying $\overline{V_j^\alpha}(\br)=0$ and
\begin{equation}
\label{meq:cdis}
\overline{V_j^\alpha(\br)V_{j'}^\beta(\br')}=V^2\delta_{\alpha\beta}\delta_{jj'}e^{-(|x-x'|+|y-y'|)/\xi_{\rm dis}},
\end{equation}
with the overline signifying disorder averaging. Note that a similar linear coupling between the superconducting order
$\psi$ and the random fields is forbidden by gauge invariance. Therefore, $\psi$ couples to the potentials only
indirectly through the constraints, Eq. (\ref{meq:constraint}).

Within a layer the Hamiltonian reads
\begin{eqnarray}
\label{meq:slh}
\nonumber
\!\!\!\!\!\!\!H_{j\mu}&=&\frac{\rho_s}{2}\sum_\br\Big[\left|\bnabla\psi_{j\mu}\right|^2 +\lambda|\bnabla \bPhi_{j\mu}|^2
+g|\bPhi_{j\mu}|^2\\
\!\!\!\!\!\!\!&&
+(\tilde{\bm V}^\dagger_{j\mu}\bPhi_{j\mu}+{\rm c.c.}) \Big],
\end{eqnarray}
where $\bnabla$ is the discrete gradient, $\lambda\rhos$ is the CDW stiffness and $g\rhos$ is the effective CDW mass reflecting
the energetic penalty for CDW ordering. The presence of such a penalty ensures that superconductivity prevails over the CDW order
at $T=0$, at least in the disorder-free regions. Finally, $\tilde {\bm V}_{j\mu}=(\tilde V^1_j + i\tilde V^2_j, \tilde V^3_j + i\tilde V^4_j)^T$
is the disorder potential on the CuO$_2$ layers, which we model by Gaussian random fields with zero mean and
\begin{equation}
\label{meq:planedis}
\overline{\tilde V_{j\mu}^\alpha(\br)\tilde V_{j'\mu'}^\beta(\br')}=\tilde V^2\delta_{\alpha\beta}\delta_{jj'}\delta_{\mu\mu'}\delta_{\br\br'}.
\end{equation}

\section{Monte Carlo Results}

We have used Monte Carlo simulations to study the model on an $L\times L \times L_z$ lattice, and in particular to calculate
the CDW structure factor
\begin{eqnarray}
\label{meq:Sdef}
\nonumber
S_\alpha(\bq,l)&=&\frac{2}{L^2 L_z}\sum_{\br\br'}\sum_{jj'}\sum_{\mu\mu'} e^{-i\left[\bq\cdot \left(\br-\br'\right)
+ 2\pi \left(j-j'+\frac{\mu-\mu'}{3}\right)l\right]}\\
&&\times \langle\Phi^\alpha_{j\mu}(\br)\Phi^{*\alpha}_{j'\mu'}(\br')\rangle,
\end{eqnarray}
where the averaging is over both thermal fluctuations and disorder realizations. The in-plane wave-vector $\bq$ in Eq. (\ref{meq:Sdef})
is measured relative to $\bQ_{a,b}$, and the expression for $S_\alpha$ reflects the fact that in YBCO the CuO$_2$ planes within a
bilayer are separated by approximately 1/3 of the $c$ axis lattice constant.

\begin{figure}[t!!!]
\centering
\includegraphics[width=0.98\linewidth,clip=true]{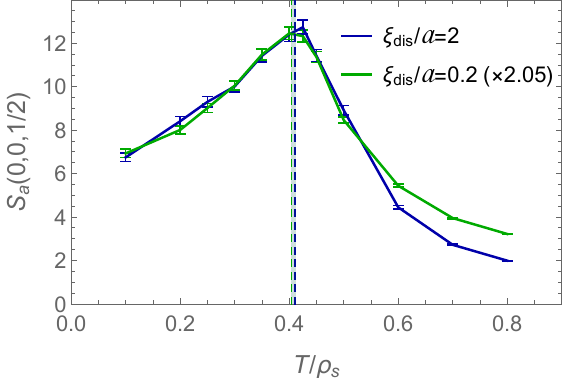}
\caption{The temperature dependence of the peak amplitude $S_a(0,0,1/2)$ for a $32^3$ system with
$V^2=0.035$, $\xi_{\rm dis}=2$ (blue) and $V^2=0.35$, $\xi_{\rm dis}=0.2$ (green). In both cases $\tilde V^2=0$.
The data for the system with $\xi_{\rm dis}=0.2$ is scaled by a factor 2.05.
The dashed vertical lines denote $T_c$ for the two cases.}
\label{fig:S-compare}
\end{figure}
\vspace{1mm}
\begin{figure}[h!!!]
\centering
\includegraphics[width=0.96\linewidth,clip=true]{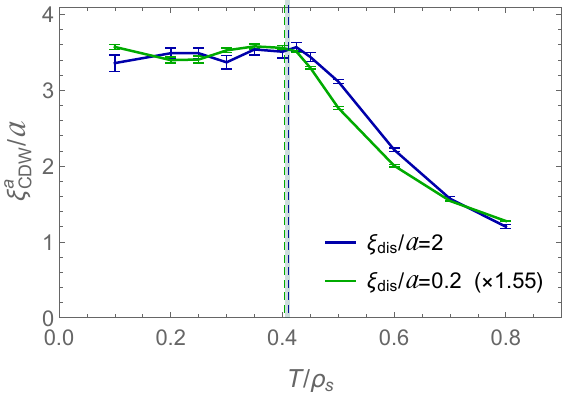}
\caption{The temperature dependence of the CDW correlation length for the same systems as in Fig. \ref{fig:S-compare}.
The data for the system with $\xi_{\rm dis}=0.2$ is scaled by a factor 1.55.}
\label{fig:xi-compare}
\end{figure}

\subsection{Correlated chain layer potential}

We begin by examining the consequences of varying the correlation length of the chain layer potential. Specifically,
we compare the case with $\xi_{\rm dis}=2$, which corresponds to the reported correlation length in the ordered ortho
samples of Ref. \cite{Achkar-disorder}, and a system with $\xi_{\rm dis}=0.2$, which essentially lacks any spatial
disorder correlations. The correlated potential is generated from a Gaussian random field
$v$ via $\cF^{-1}[\cF(v)\sqrt{\cF(e^{-(|x|+|y|)/\xi_{\rm dis}})}]$, where $\cF$ denotes the Fourier transform.
Each data point was typically averaged over 540 disorder realizations and throughout the simulations
we have set $\lambda=1$, $g=1$, $\tU=0.85$, $U=0.12$, $\tJ=0.15$, and $J=0.015$.

\begin{figure}[t!!!]
\begin{center}
\begin{tabular}{ll}
\includegraphics[width = 0.251\textwidth,clip=true]{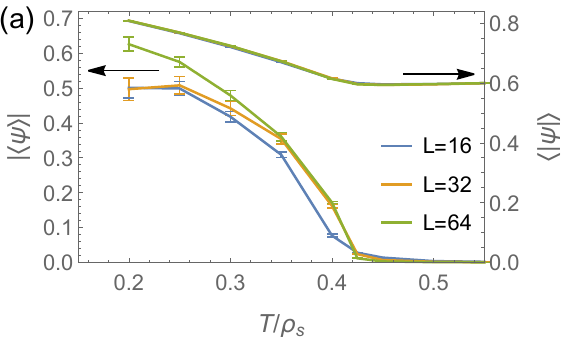} &
\includegraphics[width = 0.214\textwidth,clip=true]{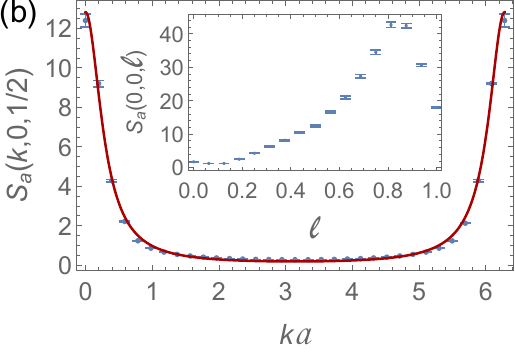}
\end{tabular}
\caption{(a) The average superconducting order parameter and its average magnitude for a system of size $L\times L\times 16$ with
$V^2=0.035$, $\xi_{\rm dis}=2$ and $\tilde V^2=0$. (b) A Lorentzian fit of $S_a(k,0,1/2)$ for a $32^3$ system with similar parameters
at $T=0.4$. The inset depicts the $l$ dependence of $S_a$.}
\label{fig:Tc-xi-fit}
\end{center}
\end{figure}

Figure \ref{fig:S-compare} depicts the temperature dependence of the structure factor peak at $\bQ_a$ and $l=1/2$.
The peak amplitude increases as the temperature is lowered, attains a maximum at $T_c$ (which we identify from the onset
of superconducting order, see Fig. \ref{fig:Tc-xi-fit}a) and then decreases below $T_c$. It remains finite down to zero
temperature due to nucleation of CDW in domains that are dominated by the coupling to the chains potential, while
superconductivity sets in within the intervening regions.

We find that a correlated chain potential is more effective in establishing coherent CDW order. For example, reducing
$\xi_{\rm dis}$ from 2 to 0.2 while keeping all other parameters fixed causes $S_a$ to shrink by a factor of approximately 4
(concomitantly, $T_c$ increases by only 10\%). Hence, to maintain the experimentally observed 2:1 ratio
between $S_a$ of the two systems \cite{Achkar-disorder} we need to increase the variance of the potential by tenfold
when we eliminate its spatial correlations.
This change also leads to a similar temperature dependence of $S_a$ for the two systems, as shown by Fig. \ref{fig:S-compare}.
Such behavior aligns with the experimental findings \cite{Achkar-disorder}.

We extract the in-plane CDW correlation length by fitting $S_a(k,0,1/2)$ to a Lorentzian
$c+b/(k^2+\xi_{\rm CDW}^{-2})+b/((2\pi-k)^2+\xi_{\rm CDW}^{-2})$, as exemplified in Fig. \ref{fig:Tc-xi-fit}b.
The results, presented in Fig. \ref{fig:xi-compare} show that, similar to the amplitude of the structure factor, the CDW correlation
length retains its temperature dependence upon reducing $\xi_{\rm dis}$ and rescaling of $V^2$. It remains essentially constant
up to $T_c$ and then diminishes with increasing temperature. Nevertheless, we find that complete disordering of the chain layer
reduces $\xi_{\rm CDW}$ by one third, which is significantly larger than the observed reduction \cite{Achkar-disorder}.

To conclude this section, we note that $S_a$ exhibits an asymmetric peak around $l=0.8$,
as shown in the inset of Fig. \ref{fig:Tc-xi-fit}b. This arises from the form factor in Eq. (\ref{meq:Sdef}).
We have checked that $S_a$ presents the same qualitative behavior outlined above also at $l=0.8$ or when
only the diagonal terms ($\mu=\mu'$) are kept in Eq. (\ref{meq:Sdef}).

\subsection{Gaussian disorder on the CuO$_2$ planes}

\begin{figure}[t!!!]
\centering
\includegraphics[width=0.98\linewidth,clip=true]{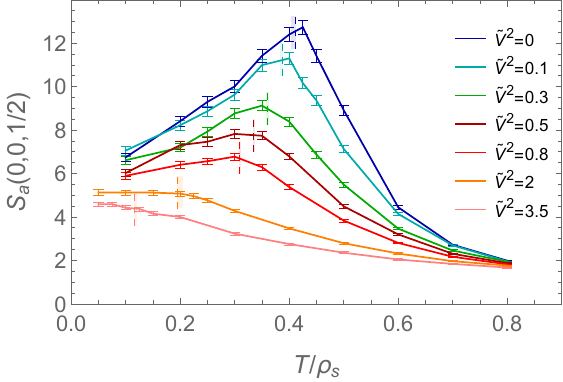}
\caption{The temperature dependence of the peak amplitude $S_a(0,0,1/2)$ for $32^3$ systems with
$V^2=0.035$, $\xi_{\rm dis}=2$ and various values of $\tilde V^2$.
For each case the corresponding $T_c$ is denoted by a dashed vertical line.}
\label{fig:S-disorder}
\end{figure}

Next, we fix $\xi_{\rm dis}=2$, $V^2=0.035$ for the potential in the chain layers and consider the effects of local Gaussian
disorder on the CuO$_2$ planes. Figure \ref{fig:S-disorder} depicts the temperature evolution of the structure factor for
various values of the in-plane disorder variance $\tilde V^2$. Evidently, when the disorder is not too strong $S_a$
exhibits a maximum, which coincides (within numerical errors) with $T_c$. The maximum moves to lower temperatures and
becomes less pronounced as $\tilde V^2$ increases, until it eventually disappears. Instead, we find at $T_c$ a break
in the slope of $S_a$ or, for the most disordered systems, a smooth decrease of the structure factor with temperature.

\begin{figure}[h!!!]
\centering
\includegraphics[height=0.66\linewidth,clip=true]{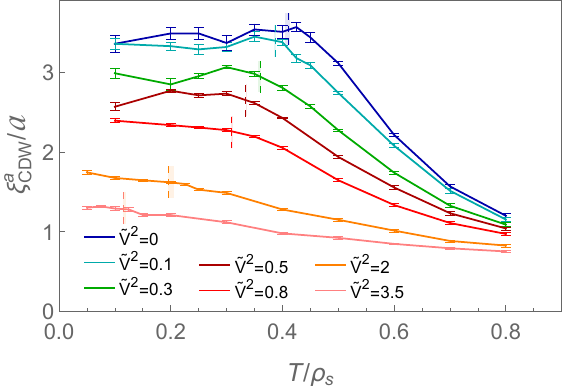}
\caption{The temperature dependence of the CDW correlation length for the same systems as in Fig. \ref{fig:S-disorder}.}
\label{fig:xi-disorder}
\end{figure}

\begin{figure}[t!!!]
\centering
\includegraphics[height=0.665\linewidth,clip=true]{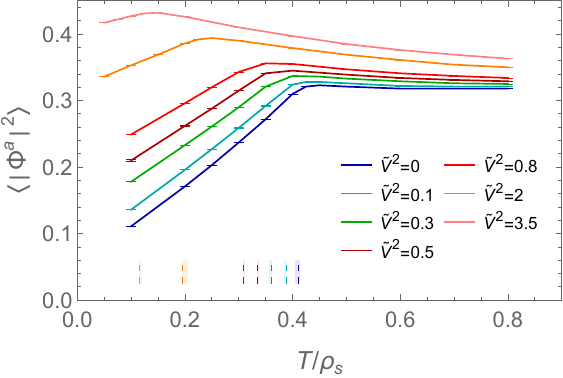}
\caption{The temperature evolution of the squared magnitude of the CDW field $\Phi^a$
for the same systems as in Fig. \ref{fig:S-disorder}.}
\label{fig:phi}
\end{figure}

Insights into the reasons behind this behavior of $S_a$ can be gleaned from the CDW correlation length, as presented
in Fig. \ref{fig:xi-disorder}, and the average (squared) magnitude of the CDW field $\Phi^a$, as shown in Fig. \ref{fig:phi}.
For weak disorder the correlation length increases with decreasing temperature and then remains
approximately constant below $T_c$. At the same time, the magnitude of the CDW order parameter
is approximately constant above $T_c$ and then decreases below $T_c$. Hence, the increase of the CDW structure
factor above $T_c$ is due to the establishment of longer range CDW coherence while its decrease below $T_c$ is driven
by the decrease in the average magnitude of the CDW order parameter due to its competition with superconductivity.
For strong disorder the local magnitude of the CDW order is enhanced by the interaction with the potential on the
CuO$_2$ layers, but its correlation length is smaller. Both change more gradually than in the cleaner
systems, leading to a smoother evolution of the structure factor. The overall decrease in the magnitude of the
structure factor with increasing disorder is largely due to the reduction in the correlation length.

We have already noted that gauge invariance precludes direct linear coupling between the disorder and the superconducting
order. Consequently, the reduction in $T_c$ with increasing $\tilde V^2$ can result from two effects. First, the suppression
of the magnitude of the superconducting order by the competition with the locally enhanced CDW. Secondly, by
phase disordering due to the same competition with the disordered CDW fields. Figure \ref{fig:phi} together with the constraint
Eq. (\ref{meq:constraint}) show that in the weak disorder regime the superconducting amplitude changes relative little
above $T_c$ as a function of the disorder strength. Hence, the reduction of $T_c$ in this regime is mostly driven by
phase fluctuations. For strong disorder the superconducting amplitude is more significantly suppressed and this effect
also contributes to the decrease in $T_c$.

\section{Large $N$ analysis}

Analytical progress in analyzing the problem can be made by enlarging the number of real components of the vector of order parameters
from 6 to $N\gg 1$. The superconducting order is described by the real fields $n_{j\mu}^\alpha(\br)$, with $\alpha=1,\cdots, N/3$,
while the CDW fields $\Phi^{a,b}$ are described by $n_{j\mu}^\alpha(\br)$, with $\alpha=N/3+1,\cdots, 2N/3$, and $\alpha=2N/3+1,\cdots, N$,
respectively. The Hamiltonian becomes
\begin{eqnarray}
  \label{eq:HVN}
  \nonumber
  H&=&\sum_{\alpha=1}^N\sum_{j\mu}\int\!\dr\left[\frac{\lambda_\alpha}{2}(\bnabla n^{\alpha }_{j\mu})^2+\frac{g_\alpha}{2}(n_{j\mu}^{\alpha})^2
  +\tilde\cV_{j\mu}^\alpha n_{j\mu}^{\alpha}\right]\\
  \nonumber
  &&+\sum_{\alpha=1}^N\sum_j\int \! \dr\Big[
  \tW_\alpha n^{\alpha }_{j0}n^{\alpha }_{j1}+W_\alpha n^{\alpha }_{j1}n^{\alpha }_{j+1,0} \\
  &&+ \cV_j^{\alpha}\left( n^\alpha_{j1}+n^{\alpha }_{j+1,0}\right)\Big],
\end{eqnarray}
where we have already set $\rhos=1$. In order to be able to carry out the disorder averaging we assume that {\it both} the chain layer
potentials $\cV_j^\alpha$ and the CuO$_2$ layer potentials $\tilde\cV_{j\mu}^\alpha$ are independent local random Gaussian fields with
respective variances $V_\alpha^2/N$ and $\tilde V_\alpha^2/N$, scaled to ensure $N$-independent results as $N\rightarrow\infty$.
In Eq. (\ref{eq:HVN}), $g_\alpha=0$, $W_\alpha=-J$,
$\tW_\alpha=-\tJ$, $V_\alpha=0$ and $\tilde V_\alpha=0$ for the superconducting components, while for the CDW components $g_\alpha=g$, $W_\alpha=U$,
$\tW_\alpha=\tU$, $V_\alpha=V$ and $\tilde V_\alpha=\tilde V$.

Taking into account the constraints of the NLSM, the partition function reads
\begin{equation}
  \label{eq:FV}
  Z=e^{-N\beta F}=  \int\D n\,\prod_{j\mu}\delta\left[
    \sum_\alpha (n_{j\mu}^\alpha)^2-1\right]e^{-S},
\end{equation}
with
\begin{eqnarray}
\nonumber
\!\!\!\!\!S=&&\,N\beta H\\
&&+ \int d^2rd^2r'\!\!\sum_{\alpha\beta jj' \mu\mu'}\!\!K^{\alpha\beta} _{jj' \mu\mu'}(\br,\br')
  n_{j\mu}^{\alpha }(\br)n^{\beta }_{j'\mu'}(\br'),
\end{eqnarray}
where the scaling $N\beta$ is used to facilitate a sensible large-$N$ limit.
The source term $K$ is introduced for the purpose of evaluating the Green's functions via
\begin{equation}
  \label{eq:GviaK}
  {\cal G}^{\alpha\beta}_{jj' \mu\mu'}(\br,\br')\equiv\overline{\braket{n^{\alpha}_{j\mu}(\br)n^{\beta}_{j'\mu'}(\br')}}
    =\left.\frac{\delta\,N\beta\overline{F}}{\delta K^{\alpha\beta}_{jj' \mu\mu'}(\br,\br')}
      \right|_{K=0}.
\end{equation}

\begin{figure}[t!!!]
\vspace{-0.1cm}
\centering
\includegraphics[height=0.655\linewidth,clip=true]{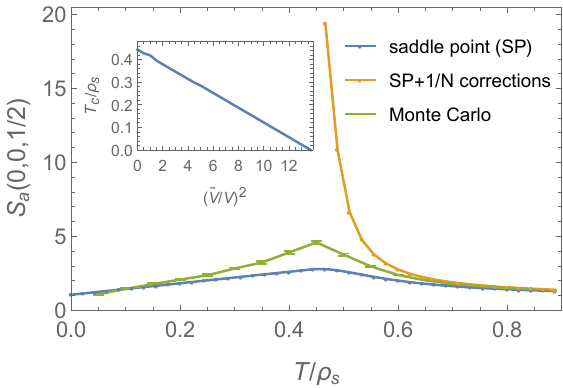}
\caption{Comparing the saddle-point results, with and without $1/N$-corrections, to Monte Carlo calculations of
$S_a(0,0,1/2)$ for a $32^2\times 16$ systems with Gaussian $V^2=0.1$ chain-layer potential and $\tilde V=0$.
The inset depicts the saddle-point $T_c$ as a function of $\tilde V^2/V^2$.}
\label{fig:S-sp-MC-v0}
\end{figure}

\begin{figure}[t!!!]
\centering
\includegraphics[height=0.64\linewidth,clip=true]{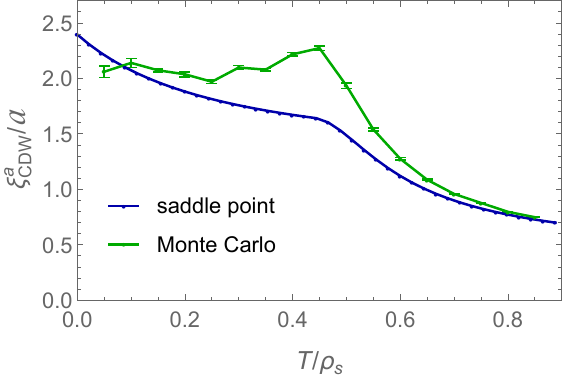}
\caption{The CDW correlation length for the system presented in Fig. \ref{fig:S-sp-MC-v0}, calculated using the
saddle point approximation and Monte Carlo simulations.}
\label{fig:xi-sp-MC-v0}
\end{figure}

To calculate $\overline{F}$ we consider $R$ replicas of the model and use the replica trick
$\overline{F}=\lim_{R\rightarrow 0}F_R/R$, where
\begin{eqnarray}
\label{eq:Fm}
\nonumber
e^{-N\beta F_R}&=&\int \D n\, \D\cV\, \D\tilde\cV\, \D\sigma \exp\Bigg\{-\sum_{a=1}^R S\left(n^{\alpha a}_{j\mu}\right) \\
\nonumber
&& +\int d^2r \Bigg[ i\frac{N\beta}{2}\sum_{j\mu}\sum_{a=1}^R\sigma^a_{j\mu}\Bigg(1-\sum_\alpha(n_{j\mu}^{\alpha a})^2\Bigg) \\
&&  -N\sum_{j \alpha} \left( \frac{|\cV_j^\alpha|^2}{2V_\alpha^2}+\sum_\mu \frac{|\tilde \cV_{j\mu}^\alpha|^2}{2\tilde V_\alpha^2}\right)\Bigg]\Bigg\}.
\end{eqnarray}
Here, $a$ is the replica index and we have incorporated the constraints by an integral over the Lagrange multiplier fields $\sigma_{j\mu}^a$.
Carrying out the disorder integrals we find $e^{-N\beta F_R}=\int\D n\D\sigma e^{-\tilde S_R}$, given in terms of
\begin{eqnarray}
\label{eq:Stilde}
\nonumber
\tilde S_R &=& \frac{1}{2}\int d^2rd^2r' \sum_{\alpha\beta jj' ab} \left[\begin{array}{cc}
     n_{j0}^{\alpha a}(\br) & n_{j1}^{\alpha a}(\br)\end{array}\right]\\
\nonumber
&&\times \left[\left(\hat G^{-1}+2\hat K\right)_{jj'}^{\alpha\beta ab}\!\!(\br,\br')\right]
  \left[\begin{array}{c} n_{j'0}^{\beta b}(\br') \\ n_{j'1}^{\beta b}(\br')\end{array}\right] \\
&&-i\frac{N\beta}{2}\int d^2r \sum_{j\mu a}\sigma_{j\mu}^a ,
\end{eqnarray}
where a hat denotes a $2\times 2$ matrix with indices $\mu,\mu'$. We also define
$\hat K^{\alpha\beta ab }_{\mu\mu'jj'}(\br,\br')=\delta_{ab}K^{\alpha\beta}_{\mu\mu' jj'}(\br,\br')$, and
\begin{equation}
  \label{eq:Ginv}
  (\hat G^{-1})^{\alpha\beta ab}_{ jj'}(\br,\br') = N\beta\left[\delta_{ab}\hat L^{\alpha a}_{ jj'}
      -\hat M^\alpha_{ jj'}\right]\delta_{\alpha\beta}\delta(\br-\br').
\end{equation}
The latter contains the interaction and disorder parts
\begin{eqnarray}
\label{eq:Lmat}
\!\!\!\!\!\!\!\!\!&&\hat{L}_{ jj'}^{\alpha a}=\left[ \begin{array}{cc}
\cL_{jj'0}^{\alpha a}  & \tW_\alpha \delta_{j'j} +W_\alpha \delta_{j',j-1} \\
\tW_\alpha \delta_{j'j} +W_\alpha \delta_{j',j+1}  & \cL_{jj'1}^{\alpha a} \end{array} \right],\\
\!\!\!\!\!\!\!\!\!&&\hat{M}^\alpha_{ jj'}= \beta V_\alpha^2 \left[ \begin{array}{cc}
(1+\upsilon_\alpha^2)\delta_{jj'} & \delta_{j' j-1}  \\
\delta_{j' j+1}  & (1+\upsilon_\alpha^2)\delta_{jj'}
\end{array} \right],
\end{eqnarray}
involving
\begin{eqnarray}
\label{eq:Lop}
\cL_{jj'\mu}^{\alpha a}&=&\left[-\lambda_\alpha\nabla^2+g_\alpha+i\sigma_{j\mu}^a(\br)\right]\delta_{jj'} ,\\
\upsilon_\alpha^2&=&{\tilde V_\alpha}^2/V_\alpha^2.
\end{eqnarray}

Next, we integrate out the $n$ fields. To allow for spontaneous symmetry breaking
in the superconducting channel, which amounts to $(n^1,n^2)=O(1)$, we denote $\psi_{j\mu}^a=n^{1a}_{j\mu}+in^{2a}_{j\mu}$
and avoid integrating over it (consequently, henceforth, $\alpha=3,\cdots,N$). The result is
\begin{equation}
\label{eq:Fm2}
e^{-N\beta F_R} =\int\D\sigma^a\D\psi^a e^{-S_R},
\end{equation}
where
\begin{eqnarray}
\label{eq:Sm}
\nonumber
S_R&=&\frac{1}{2}\Tr\ln(G^{-1}+2K) \\
\nonumber
&& +\frac{N\beta}{2}\int d^2r\Bigg\{\sum_{j\mu a}\left[|\bnabla\psi_{j\mu}^a|^2+i\sigma_{j\mu}^a(|\psi_{j\mu}^a|^2-1)\right] \\
&&-\sum_{j a}\left(\tJ\psi_{j0}^{a*}\psi_{j1}^a+J\psi_{j1}^{a*}\psi_{j+1,0}^a+{\rm c.c.}\right)\Bigg\} .
\end{eqnarray}

\begin{figure}[t!!!]
\vspace{-0.07cm}
\centering
\includegraphics[height=0.64\linewidth,clip=true]{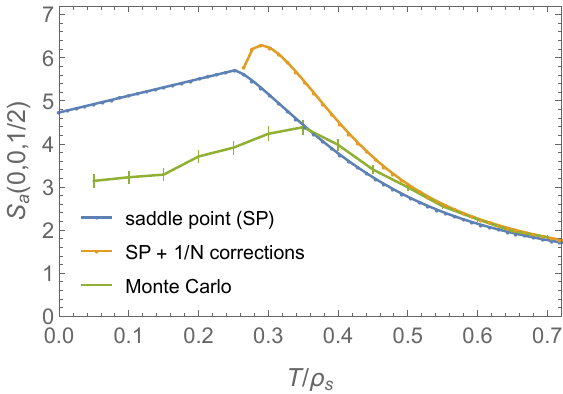}
\caption{Comparing the saddle-point results, with and without $1/N$-corrections, to Monte Carlo calculations of
$S_a(0,0,1/2)$ for a $32^2\times 16$ systems with Gaussian $V^2=0.1$ chain-layer potential and Gaussian $\tilde V^2=0.6$
disorder on the CuO$_2$ planes.}
\label{fig:S-sp-MC-v6}
\end{figure}

\subsection{Saddle-point approximation}

The integrals over $\sigma^a_{j\mu}$ and $\psi_{j\mu}^a$ in Eq. (\ref{eq:Fm2}) are to be calculated using a saddle-point approximation,
which is justified in the limit $N\to\infty$. Within this approximation, $N\beta F_R= S_R$, where $S_R$ is evaluated using its saddle-point
configurations. To evaluate the Green's function $\cG^{\alpha\beta}_{ \mu\mu' jj'}(\br,\br')$ it is therefore necessary to know
$S_R$ to first order in $K$. One can check that it is sufficient for this purpose to use the saddle point configurations of
$S_R^0\equiv S_R|_{K=0}$, as this entails only ${\cal O}(K^2)$ corrections. Varying $S_R^0$ with respect to $\sigma_{j\mu}^a(\br)$
and $\psi_{j\mu}^{a*}(\br)$ yields the saddle point equations
\begin{eqnarray}
\label{eq:SPV1}
\!\!\!\!\!\sum_\alpha G^{\alpha\alpha aa}_{jj\mu\mu}(\br,\br)&=&1-|\psi_{j\mu}^a|^2 , \\
\label{eq:SPV2}
\nonumber
\!\!\!\!\!\left(-\bnabla^2+i\sigma^a_{\mu j}\right)\psi_{\mu j}^a&=&\tJ\left(\delta_{\mu0}\psi_{1j}^a +\delta_{\mu1}\psi_{0j}^a\right) \\
\!\!\!\!\!&&+ J\left(\delta_{\mu0}\psi_{1j-1}^a +\delta_{\mu1}\psi_{0j+1}^a\right).
\end{eqnarray}
Since $G^{-1}$ is diagonal
in $\alpha,\beta$ and symmetric in $a,b$, $\br,\br'$, and under exchange of both $j,j'$ and $\mu,\mu'$ so is
$G^{\alpha\beta ab}_{jj'\mu\mu'} = \delta_{\alpha\beta}G_{jj'\mu\mu'}^{\alpha ab}$. Thus, it follows from
Eqs. (\ref{eq:GviaK}) and (\ref{eq:Sm}) that
\begin{equation}
  \label{eq:Glim}
  \hat{\cG}_{jj'}^{\alpha\beta}(\br,\br')\equiv \delta_{\alpha\beta}\hat{G}_{jj'}^\alpha(\br,\br')
  =\delta_{\alpha\beta}\lim_{m\to 0}\frac{1}{m}
  \sum_a \hat {G}_{jj'}^{\alpha aa}(\br,\br'),
\end{equation}
where $\hat G$ is evaluated using the saddle point solutions.

\begin{figure}[t!!!]
\centering
\includegraphics[height=0.64\linewidth,clip=true]{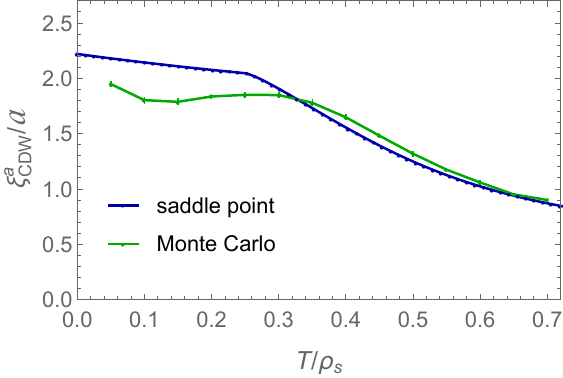}
\caption{The CDW correlation length for the system presented in Fig. \ref{fig:S-sp-MC-v6}, calculated using the
saddle point approximation and Monte carlo simulations.}
\label{fig:xi-sp-MC-v6}
\end{figure}

We will calculate the $2\times2$ correlation matrix $\hat{G}_{jj'}^{\alpha ab}$ by assuming a replica-symmetric
solution of the saddle-point equations, which is also independent of $j$, $\mu$ and $\br$, \ie, $\psi_{\mu j}^a=\psi$,
$\sigma_{\mu j}^a(\br)=-i\sigma$. Under this assumption the interaction part, Eq. (\ref{eq:Lmat}), is also replica symmetric
$\hat {L}_{jj'}^{\alpha a}=\hat {L}_{jj'}^\alpha$ and $\hat {G}_{jj'}^{\alpha ab}$ is determined from
\begin{eqnarray}
\label{eq:GGinv}
\nonumber
&&N\beta\sum_{cl}\int d^2\tilde r\left[\delta_{ac}\hat{L}^\alpha_{jl} -\hat{M}^\alpha_{jl}\right]\delta(\br-\tilde\br)
 \hat{G}_{lj'}^{\alpha cb}(\tilde\br,\br') \\
&&=\hat I\delta_{jj'}\delta_{ab}\delta(\br-\br').
\end{eqnarray}
Expanding
\begin{equation}
  \label{eq:Geigen}
  \hat {G}_{ jj'}^{\alpha ab}(\br,\br')=\frac{2}{L^2 L_z}\sum_{\bk,k_z}\hat{G}^{\alpha ab}(k)
  e^{i[\bk\cdot(\br-\br')+k_z(j-j')]},
\end{equation}
we find that Eq. (\ref{eq:GGinv}) is solved by $\hat {G}^{\alpha ab}(k)=\hat A^\alpha(k)\delta_{ab}+\hat B^\alpha(k)$, where,
henceforth, $k\equiv(\bk,k_z)$.
This leads by Eq. (\ref{eq:Glim}) to the correlation matrix
\begin{equation}
\label{eq:Gsol}
\hat {G}^\alpha(k)=\hat A^\alpha(k)+\hat B^\alpha(k),
\end{equation}
with
\begin{eqnarray}
\label{eq:solA}
\nonumber
\!\!\!\!\!\!\!\!\!\hat A^\alpha(k)&=&\frac{1}{N\beta}\frac{1}{\epsilon_\alpha^2(\bk)-\epsilon_{\alpha\perp}^2(k_z)} \\
\!\!\!\!\!\!\!\!\!&&\times
\left[\begin{array}{cc}
\epsilon_\alpha(\bk) &  -\tW_\alpha - W_\alpha e^{-ik_z} \\
-\tW_\alpha - W_\alpha e^{ik_z}  & \epsilon_\alpha(\bk)
\end{array}\right] ,
\end{eqnarray}
and where the components of $\hat B_\alpha$ are given by
\begin{eqnarray}
\nonumber
\!\!\!\!\!(B^\alpha)_{00}(k)&=&(B^\alpha)_{11}(k)\\
\nonumber
&=&\frac{1}{N}\frac{V_\alpha^2}{\left[\epsilon_\alpha(\bk)^2-\epsilon_{\alpha\perp}^2(k_z)\right]^2} \\
\nonumber
&&\times\Big\{\left(1+\upsilon_\alpha^2\right)\left[\epsilon_\alpha(\bk)^2+\epsilon_{\alpha\perp}^2(k_z)\right]\\
&&-2\epsilon_\alpha(\bk)\left( W_\alpha+\tW_\alpha\cos k_z\right)\Big\},\\
\nonumber
\!\!\!\!\!(B^\alpha)_{01}(k)&=&(B^\alpha)_{10}^*(k)\\
\nonumber
&=&\frac{1}{N}\frac{V_\alpha^2e^{-ik_z}}{\left[\epsilon_\alpha(\bk)^2-\epsilon_{\alpha\perp}^2(k_z)\right]^2}\\
\nonumber
&&\times\Big\{\left[W_\alpha-\epsilon_\alpha(\bk)+\tW_\alpha e^{ik_z}\right]^2\\
&&-2\upsilon_\alpha^2\epsilon_\alpha(\bk)\left(W_\alpha+\tW_\alpha e^{ik_z}\right)\Big\}.
\end{eqnarray}
Here,
\begin{eqnarray}
\label{eq:epskdef}
\nonumber
&&\epsilon_\alpha(\bk)=\left\{\begin{array}{cc}\lambda_\alpha (k_x^2+k_y^2)+g_\alpha+\sigma & {\!\!\!\!\!\!\!\rm continuum} \\
2\lambda_\alpha (2-\cos k_x-\cos k_y)+g_\alpha+\sigma & {\!\!\!\rm lattice}
\end{array}\right. , \\
\\
&&\epsilon_{\alpha\perp}(k_z)=\sqrt{W_\alpha^2+\tW_\alpha^2+2W_\alpha\tW_\alpha\cos k_z}.
\end{eqnarray}

Under our ansatz, the saddle point equation (\ref{eq:SPV2}) becomes $\psi(\sigma-J-\tJ)=0$.
It is either solved by $\psi=0$ and $\sigma$ that is then determined by Eq. (\ref{eq:SPV1}), or by $\sigma=J+\tJ$
and $\psi$ that is given by Eq. (\ref{eq:SPV1}). The first solution holds at high temperatures. When $T$ decreases,
$\sigma$ decreases with it, reaching $J+\tJ$ at $T=T_c$ and sticking at this value for lower temperatures where
$\psi\neq 0$. As the inset of Fig. \ref{fig:S-sp-MC-v0} shows, we find that the saddle point estimate of $T_c$
vanishes linearly with $\tilde V^2/V^2$.

In Figs. \ref{fig:S-sp-MC-v0}-\ref{fig:xi-sp-MC-v6} we compare the saddle point predictions for the temperature
evolution of the structure factor and the CDW correlation length with Monte Carlo results for the $N=6$ model.
To this end, we note that $S_a(\bk,k_z)=4G^{\alpha}_{00}(\bk,kz)+4{\rm Re}[G^{\alpha}_{01}(\bk,kz)\exp(ik_z/3)]$,
where $\hat G^{\alpha}$ is the correlation matrix for one of the CDW components. Furthermore, the scaling of $T$ and $V^2$,
used in defining the large-$N$ model, implies that they correspond to $T/N$ and $V^2/N$ in the Monte Carlo simulations.
Our results show that for the case $\tilde V^2=0$, the saddle point approximation yields a reasonable estimate for
$T_c$ as well as for the high and low temperature behavior of $S_a$. However, it underestimates $S_a$ in the
vicinity of $T_c$. The same statements hold for $\xi_{\rm CDW}$. The picture changes for $\tilde V^2>0$,
where the saddle point approximation underestimates $T_c$ and overestimates both $S_a$ and $\xi_{\rm CDW}$
at low temperatures.

\subsection{$1/N$ corrections}

Next, we consider $1/N$ corrections to the saddle-point \cite{PodolskyHiggs}. We do so for the symmetric phase $\psi=0$
where the part of the action involving $\psi$ is identical to that of the other components with $\alpha=3,\cdots,N/3$,
and we include it in the integration over $n^\alpha$. To proceed we expand $\sigma_{j\mu}^a=-i\sigma+\eta_{j\mu}^a$
and plug it into Eq. (\ref{eq:Sm}). The result is
\begin{equation}
\label{eq:Sm2}
S_R=\frac{1}{2}\Tr\ln(G^{-1}+i\Lambda+2K)-i\frac{N\beta}{2}\int d^2 r\sum_{a\mu j}\eta_{j\mu}^a,
\end{equation}
where $G$ is evaluated at the saddle point $-i\sigma$ and
\begin{equation}
\label{eq:Lambda}
\Lambda^{\alpha\beta ab }_{jj'\mu\mu'}(\br,\br')=N\beta\eta_{j\mu}^a(\br) \delta_{\alpha\beta}\delta_{ab}
\delta_{jj'}\delta_{\mu\mu'}\delta(\br-\br').
\end{equation}
Consequently, using the symmetry properties of $G$ and the fact that it satisfies the saddle point equation (\ref{eq:SPV1})
we obtain that apart from a $K$-independent piece
\begin{equation}
\label{eq:Fm3}
N\beta F_R=\Tr(G K)-\ln\left[\int\D\lambda e^{-S_0(\Lambda)-S_{int}(\Lambda)}\right],
\end{equation}
where
\begin{equation}
\label{eq:S0Lambda}
S_0=\frac{1}{2}\sum_{ab}\sum_{\mu\mu'}\int \frac{d^3 k}{(2\pi)^3}
\eta_{\mu}^a(k){\chi^{-1}}_{\mu\mu'}^{ab}(k)\eta_{\mu'}^b(-k),
\end{equation}
with
\begin{equation}
\label{eq:xhi}
{\chi^{-1}}_{\mu\mu'}^{ab}(k)=\frac{(N\beta)^2}{2}\int\frac{d^3 q}{(2\pi)^3}\sum_\alpha G_{\mu\mu'}^{\alpha ab}(q)
G_{\mu'\mu}^{\alpha ba}(q+k).
\end{equation}
To linear order in $K$
\begin{equation}
\label{eq:SLambdaint}
S_{int}=\Tr \left[\sum_{n=1}^\infty (-i)^n(G\Lambda)^n G K -\frac{1}{2}\sum_{n=3}^\infty \frac{(-i)^n}{n}(G\Lambda)^n\right].
\end{equation}

\begin{figure}[t!!!]
\centering
\includegraphics[height=0.64\linewidth,clip=true]{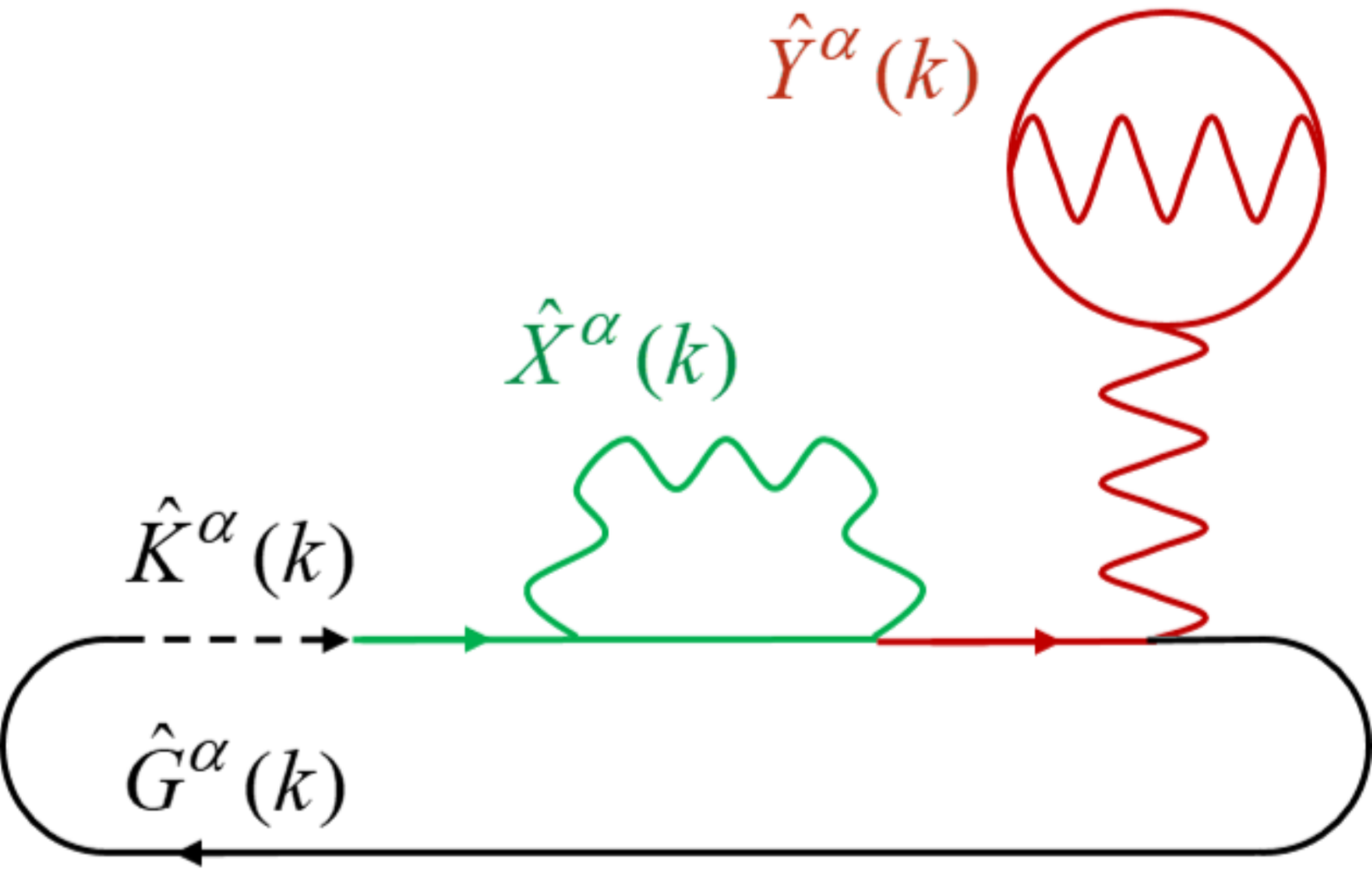}
\caption{A typical diagram that contributes to the $1/N$ correction. A broken line represents $\hat K^\alpha(k)$,
a solid line $\hat G^\alpha$ and a wavy line $\hat\chi(k)$. The full correction is the sum over such diagrams
with all possible sequences of $\hat X^\alpha(k)$ and $\hat Y^\alpha(k)$ blocks.}
\label{fig:diag}
\end{figure}
The linked cluster theorem implies
\begin{equation}
\label{eq:lct}
\ln\left[\int\D\lambda e^{-S_0-S_{int}}\right]=\ln\left[\int\D\lambda e^{-S_0}\right]+\langle e^{-S_{int}} \rangle_{\rm conn}-1,
\end{equation}
where $\langle A\rangle_{\rm conn}$ is the sum over all connected diagrams generated by averaging $A$ with respect to $S_0$.
We focus on the diagrams of the type presented in Fig. \ref{fig:diag}, which provide a $1/N$ correction to the correlation matrix.
This conclusion is based on the fact that both $\hat G^\alpha(k)$ and $\hat \chi^\alpha(k)$ scale as $1/N$, while each vertex
at the end of a $\chi$ line contributes a factor $N$. Hence, the blocks $\hat X^\alpha(k)$ and $\hat Y^\alpha(k)$ that appear
in the self energy are both of order $1/N$ ($\hat Y$ contains an additional factor of $N$ coming from the sum over the fields in the loop).
Explicitly,
\begin{eqnarray}
\label{eq:XY}
\hat X^\alpha(k)&=&-\hat G^\alpha(k) \hat F^\alpha(k) ,\\
\hat Y^\alpha(k)&=&\frac{1}{2}\hat G^\alpha(k) \hat \Gamma ,
\end{eqnarray}
where
\begin{eqnarray}
\label{eq:FGamma}
F^\alpha_{\mu\mu'}(k)&=&(N\beta)^2\int \frac{d^3 q}{(2\pi)^3}G^\alpha_{\mu\mu'}(k+q)\chi_{\mu'\mu}(q) ,\\
\nonumber
\Gamma_{\mu\mu'}&=&\delta_{\mu\mu'}(N\beta)^2\sum_{\nu_1\nu_2\nu_3}\sum_\alpha\int \frac{d^3 q}{(2\pi)^3} G^\alpha_{\nu_1\nu_3}(q) \\
&&\times F_{\nu_3\nu_2}^\alpha(q)G^\alpha_{\nu_2\nu_1}(q)\chi_{\nu_1\mu}(0).
\end{eqnarray}
In the above we have suppressed the replica indices on the various quantities, since to obtain a result that is
proportional to $R$ they must all equal each other.

Summing over the diagrams [including the zeroth order one ${\rm Tr}(GK)$], and using Eq. (\ref{eq:GviaK}),
we arrive at
\begin{equation}
\label{GN}
\hat \cG^{\alpha\beta}(k)=\delta_{\alpha\beta}\left[\hat 1-\hat X^\alpha(k) - \hat Y^\alpha(k)\right]^{-1}\hat G^\alpha(k),
\end{equation}
or
\begin{equation}
\label{invGN}
{\hat\cG}^{\alpha\beta}(k)^{-1}=\delta_{\alpha\beta}\left[ \hat G^{\alpha}(k)^{-1} +\hat F^\alpha(k) -\frac{1}{2}\hat\Gamma \right].
\end{equation}

Figure \ref{fig:S-sp-MC-v0} illustrates the impact of including the $1/N$ corrections in the case $\tilde V=0$.
Evidently, while the corrections align the analytical results more closely with the Monte Carlo simulations at high temperatures,
they significantly overestimate the CDW structure factor peak around $T_c$. In the system with $\tilde V>0$,
depicted in Fig. \ref{fig:S-sp-MC-v6}, the corrections shift the maximum of $S_a(0,0,1/2)$ to a somewhat higher
temperature, bringing it closer to the temperature at which the Monte Carlo results attain a maximum. Although the agreement
with Monte Carlo simulations also improves at high temperatures, the analytical results still overestimate
the numerical outcomes at intermediate temperatures, similar to the $\tilde V=0$ case.

%
%

\section{Conclusions}

One of the goals of the present study was to explore the extent to which the observations of Achkar et al. \cite{Achkar-disorder}
can be explained from the perspective of competing superconducting and CDW orders and their coupling to the chain-layer potential.
We have shown that one of these findings, namely, that the disordering of the oxygen ortho structure cuts the amplitude of the CDW
structure factor in half but leaves its temperature dependence unchanged, is reproduced to a good approximation by the model.
We have found that eliminating the potential correlations in the chain-layer reduces the average local amplitude of
CDW order by only a few percents. Hence, the large reduction in the model's structure factor arises from the
degradation of the CDW phase correlations. This fact is also apparent in the decrease of the CDW correlation length.
Here, however, the model departs from the experiment where the changes in correlation length were relatively minor,
especially below $T_c$.

Whereas disorder on the chain layers does not preclude the onset of long-range CDW order (with $l=1$) \cite{NLSM-bilayer},
the same is not true for disorder on the CuO$_2$ planes. Elucidating the effects of the latter on the CDW correlations at $l=1/2$
constituted the second goal of our work.
We have shown that, for sufficiently weak disorder, the structure factor attains a maximum at $T_c$,
which by itself diminishes with increasing disorder strength, predominantly due to degradation of the superconducting phase coherence.
The structure factor peak at $T_c$ arises from enhanced CDW phase correlations as the temperature is lowered toward $T_c$,
coupled with a reduction in the magnitude of the CDW order parameter once superconductivity is established below $T_c$.
As disorder is increased, the overall size of the CDW structure factor decreases.
This happens despite an increase in the magnitude
of the local CDW order due to its nucleation by the planar disorder, a fact which also contributes to the reduction of $T_c$.
At the same time however, the disorder disrupts the onset
of CDW phase coherence, causing the CDW correlation length, and with it the structure factor, to decrease. Consequently,
the temperature dependence of the structure factor becomes more monotonic; the peak at $T_c$ first transitions
into a change in slope, then eventually disappears altogether. It would be interesting to contrast these expected behaviors
with experiments where the level of disorder can be controlled by sample irradiation.

\section*{Acknowledgements}
We thank A. Carrington and S. Hayden for useful discussions.

\end{document}